\documentstyle[aps,prb,epsfig,psfig,twocolumn]{revtex}

\def\xref#1#2#3#4#5{#1:\ #2\ {\bf #3},~#4~(#5)}
\def\xbe{\begin{equation}}
\def\xee{\end{equation}}
\begin{document}
\title{Conductance statistics near the Anderson transition}
\author{Peter Marko\v{s}
}
\address{Institute of Physics, Slovak Academy of Sciences, D\'ubravsk\'a cesta 9, 842 28 bratislava, Slovakia\\
e-mail address: markos@savba.sk}
\maketitle         

\section{Introduction}
\label{markos-one}

Since the pioneering work of the Anderson \cite{Anderson-58} 
we know that disorder localizes electrons.
At zero temperature  any small amount of the disorder
localizes all electrons in the one-dimensional systems \cite{Mott-61}. In higher dimension ($d>2$ for
systems with orthogonal symmetry)
weak disorder does not destroy the metallic regime. Only when the strength of the disorder increases over
a certain specific value (critical disorder), the electron becomes localized. This phenomenon
- transition from the metallic to the insulating regime due to an increase of disorder -
is called Anderson transition
\cite{MD}.

Scaling theory of the Anderson transition uses  the conductance $g$ \cite{LT} as the order parameter. It is
supposed \cite{Wegner-76,AALR,GLK} 
that the system size dependence of the conductance is determined only by the value 
of the conductance itself:
\xbe
\frac{\partial\ln g}{\partial\ln L}=\beta(g)
\label{scaling}
\xee
where $\beta(g)$ is an analytical function of $g$. $\beta$ is positive (negative) in the metallic (localized)
regimes, respectively. For dimension $d>2$ the 
function $\beta(g)$ changes its sign, being positive for $g\gg 1$ and negative in the limit $g\to-\infty$. 
There is an unstable 
fixed point $g_c$ defined as the solution of $\beta(g=g_c)=0$. 
System-size independent critical conductance  $g_c$ 
represents the critical point of the Anderson transition.

Relation (\ref{scaling}) contains no
information about the {\it microscopic} structure of the model.  
This means that the Anderson transition is universal.  The form of the $\beta$ function
is determined 
only by the physical symmetry and dimension of the system 
\cite{McKK-93}.

Soon after the formulation of the scaling theory of localization it became clear that
the conductance $g$ is not a  self-averaged quantity.  
Reproducible fluctuations of the conductance were found both in the metallic and
in the insulating regimes \cite{WW,Fowler}. 
The knowledge of the mean value $\langle g\rangle$ is therefore
not sufficient for complete description of the
transport properties.  One has to deal with the conductance distribution $P(g)$
\cite{ATAF,BS}
or, equivalently, with all cummulants of the conductance.  This is easier in
the metallic regime, where $P(g)$ is Gaussian and the conductance fluctuations are universal 
\cite{LSF,Imry,DMPK,Beenakker,Pnato,BMcK,G,PZIS,APM}
and independent on the value of the mean conductance and/or the system size.
The width of the distribution depends only on the dimension, physical symmetry
of the system \cite{LSF} and on the boundary conditions \cite{BHMMcK,RMS-01c}.
In the insulator, conductance wildly fluctuates within the ensemble of
macroscopically equivalent ensembles. It is the logarithm of the conductance 
which is distributed
normally in the the limit of large system size \cite{ATAF,BS,Pnato,MK-AP}. 

At present we have no complete analytical theory  able 
to describe   the conductance statistics at  the critical point.  
Analytical results are known only for systems of dimension $d=2+\epsilon$
($\epsilon\ll 1$) \cite{BS,AKL,CS}.
Expression for conductance cummulants \cite{AKL}
enabled to estimate the shape of the critical conductance distribution $P_c(g)$
\cite{CS}. $P_c(g)$ is system size independent and Gaussian in the neighbor
of the mean value $\langle g\rangle\sim\epsilon^{-1}$. The distribution possesses
power law tail $P_c(g)\sim g^{-1-2/\epsilon}$ for $g\to\infty$, and the delta-function
peak $\delta(g)$. 
These results can not be applied to three dimensional (3D) system
($\epsilon=1$) where $\langle g\rangle\sim 1$ \cite{MK-PM}.
Numerical simulations are  therefore crucial 
in 3D system. The first systematic numerical 
analysis of the conductance statistics in 3D was done in
\cite{MK-PM} and was followed by a series of papers  
\cite{M-EL,SO-97,M-PRL,SOK-00,SO-01,RMS-01a,SMO-02}.  In 2D systems,
critical conductance distribution was numerically 
studied in the regime of quantum Hall effect
\cite{2D} and in systems with the spin-orbit interaction
\cite{M-EL,M-JPF,RMS-01b}.

\smallskip

This paper  reviews our recent numerical data for   the conductance distribution. 
We address the question of the shape, universality and the
scaling of the critical conductance distribution.

\smallskip

In numerical calculation of the conductance
we suppose that  two opposite sites of the sample are connected to semi-infinite perfect leads
and use the multichannel Landauer formula 
\cite{Landauer}
which relates the conductance $g$ (in units $2e^2/\hbar$) to the transmission matrix  $t$:
\xbe
g={\rm Tr}~t^\dag t =\sum_{i=1}^{N_{\rm open}}\cosh^{-2}\frac{z_i}{2}
\label{Landauer}
\xee
In (\ref{Landauer})   we introduced  the variables  $z_i$, $i=1,2\dots N_{\rm open}$, 
($z_1<z_2<\dots$) which 
parametrize the eigenvalues of the matrix $t^\dag t$. 
In the limit $L_z>>L$, $z_i$ converges to $2L_z/\lambda_i$ where $\lambda_i$ is the $i$th
localization length of the quasi-one dimensional (quasi-1D) system 
\cite{Pnato,MK-PM}.
$N_{\rm open}$ is the number of open channels.

Owing to relation (\ref{Landauer}), the analysis  of the conductance can be reduced 
to the calculation of the eigenvalues of the matrix
$t^\dag t$. The general formula for $t$ was derived in \cite{Ando-91,PMcKR}.

According to the scaling theory,
critical exponents of the Anderson transition  as well as 
conductance statistics do not
depend on the microscopic details of the model.  
In numerical simulations, we consider the Anderson Hamiltonian
\xbe
{\cal H}=W\sum_n \varepsilon_n c^\dag_nc_n+\sum_{[nn']}\tau_{[nn']}c^\dag_nc_{n'}.
\label{hamiltonian}
\xee

\begin{figure}[t!]
\begin{center}
\includegraphics[width=.22\textwidth]{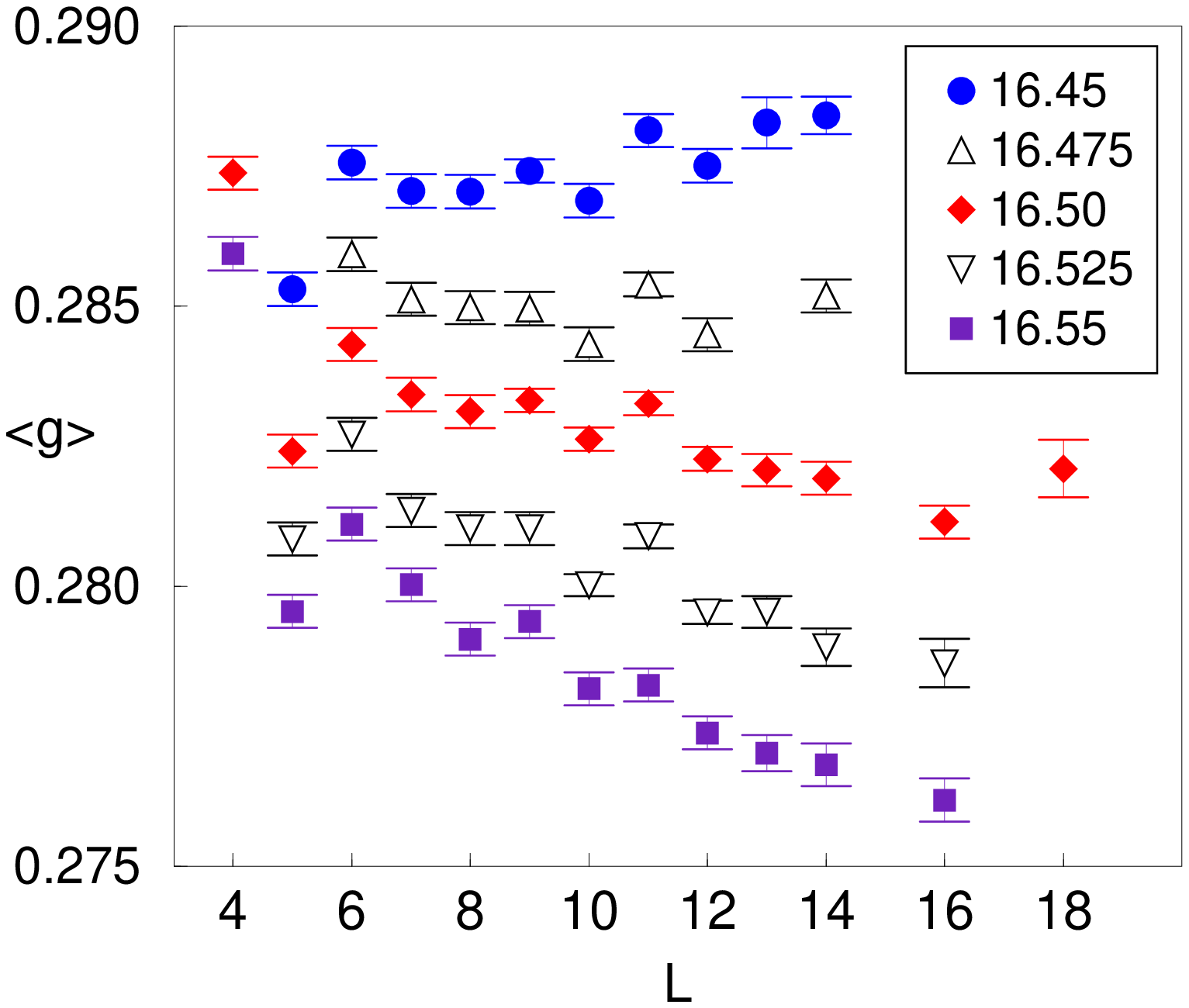}
\includegraphics[width=.22\textwidth]{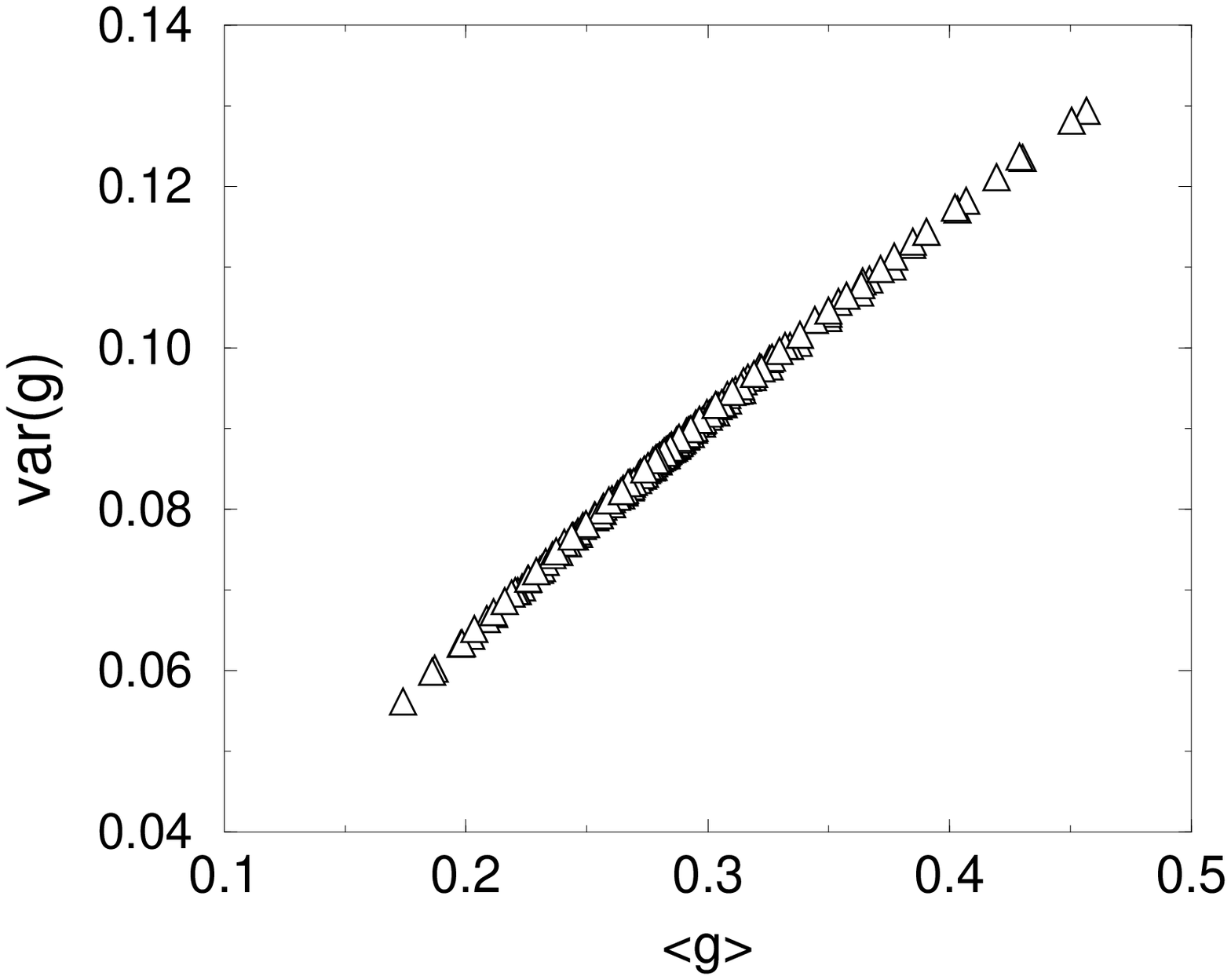}
\end{center}
\caption[]{{\it Left\,} The $L$ - dependence of the mean conductance $\langle g\rangle$ for 
different values of the disorder in the critical region. Data show that 
$16.45<W_c<16.55$. More exact estimation of the critical disorder was done in
\cite{SMO-01}. {\it Right\,}  The
unambiguous dependence var $g$ vs $\langle g\rangle$ in the critical region of the metal-insulator transition.
This agrees with the single parameter scaling theory.}
\label{m1}
\end{figure}

\noindent In (\ref{hamiltonian}) $n$ counts the  sites of the $d$-dimensional lattice and $[nn']$ 
means nearest neighbor sites. 
The hopping term $\tau$ equals to unity for orthogonal systems, unless the  anizotropy 
is considered.  When spin orbit scattering
is considered, $\tau$ is a $2\times 2$ matrix \cite{EZ,Ando-89}.  
Parameter $W$ measures the strength of the disorder. For box distribution of 
random energies $\varepsilon_n$  
($P(\varepsilon)=1$ for $|\varepsilon|\le 1/2$ and $P(\varepsilon)=0$ otherwise), 
the 3D Anderson model (\ref{hamiltonian})
exhibits Anderson transition at $W_c=16.5$.

\section{Finite size scaling}

As it was discussed in the Section \ref{markos-one}, the conductance $g$ is not 
the self-averaged quantity. To avoid  
statistical fluctuations, quasi-1D systems were introduced
\cite{PS} and  
the localization length $\gamma$  is calculated
instead of the conductance.
In the limit  $L_z/L\to\infty$  ($L_z$ and $L$ are the
length and the  width  of the system, respectively)
$\gamma$ is a self-averaged quantity which converges to its mean value.
Finite size scaling is then used \cite{McKK} for the analysis of the disorder 
and the system width dependence of $\gamma$. It is assumed that the 
variable $\Lambda=\gamma/L$ is a function of only one parameter: 
$\Lambda(L,W)=\Lambda(L/\xi(W))$.
Here $\xi=\xi(W)$ is the correlation length which   diverges in the vicinity of the critical point
$\xi(W) =|W-W_c|^{-\nu}$.
Critical exponents  $\nu$  and $s=(d-2)\nu$ characterize the critical behavior of 
the localization length and of the conductance, respectively
\cite{Wegner-76}. 

Finite size scaling analysis of the quasi-1D data enabled to test the universality of the Anderson transition
\cite{ostatni} and
provided us with the more accurate estimation of the critical exponent
$\nu\approx 1.57$ \cite{SO-99}.

\subsection{Scaling of the mean conductance}

Verification of the scaling theory of the localization requires
the proof of the universal scaling of the mean conductance and 
of the entire conductance distribution in the critical regime. 
Single parameter scaling of various mean values,
$\langle g\rangle$, $\exp\langle \ln g\rangle$, and $1/\langle g^{-1}\rangle$,
was proved numerically for the 3D Anderson model
\cite{SMO-01}. 
Statistical ensembles of more than $N_{\rm stat}\ge 10^6$ cubes
of the size from $4^3$ to $18^3$  were 
collected for various values of the disorder $W$.
Fig. \ref{m1} shows typical data for the
system size dependence of the mean conductance. In agreement with (1), 
$\langle g\rangle$
increases (decreases) with the system size in the metallic (localized) regime.
By the use of the general fitting procedure \cite{SO-99}, 
the critical disorder and the  critical exponent
$\nu\approx 1.57$ were obtained.

Data in fig. \ref{m1}  confirm that the variance, 
var $g=\langle g^2\rangle-\langle g\rangle^2$ is an unambiguous function of the
mean $\langle g\rangle$ in the critical regime.  This supports, but still
does not prove the single parameter
scaling theory.  General proof  of the theory requires 
verification of 
the single parameter scaling  of all conductance cummulants. This
is numerically
impossible since  higher cummulants  are fully determined by rare
 events with very large values of the conductance. 

\subsection{Scaling of the conductance distribution}

As higher cummulants are not treatable numerically, we test 
the scaling of the conductance
distribution by the analysis of 
the scaling of  percentiles $g_\alpha$
\cite{SMO-02}.
Percentile $g_\alpha$ is defined as
\xbe
\alpha=\int_0^{g_\alpha} P(g) {\rm d}g.
\label{percentile}
\xee
Owing to (\ref{percentile}), the probability to find $g<g_\alpha$ equals to $\alpha$.
Of course, the 
percentile $g_\alpha$ is a function of disorder 
and system size: $g_\alpha=g_\alpha(L,W)$. 
Single parameter scaling of percentiles has been proved
for several values of $\alpha$ \cite{SMO-02}. 

\begin{figure}[t!]
\begin{center}
\includegraphics[width=.22\textwidth]{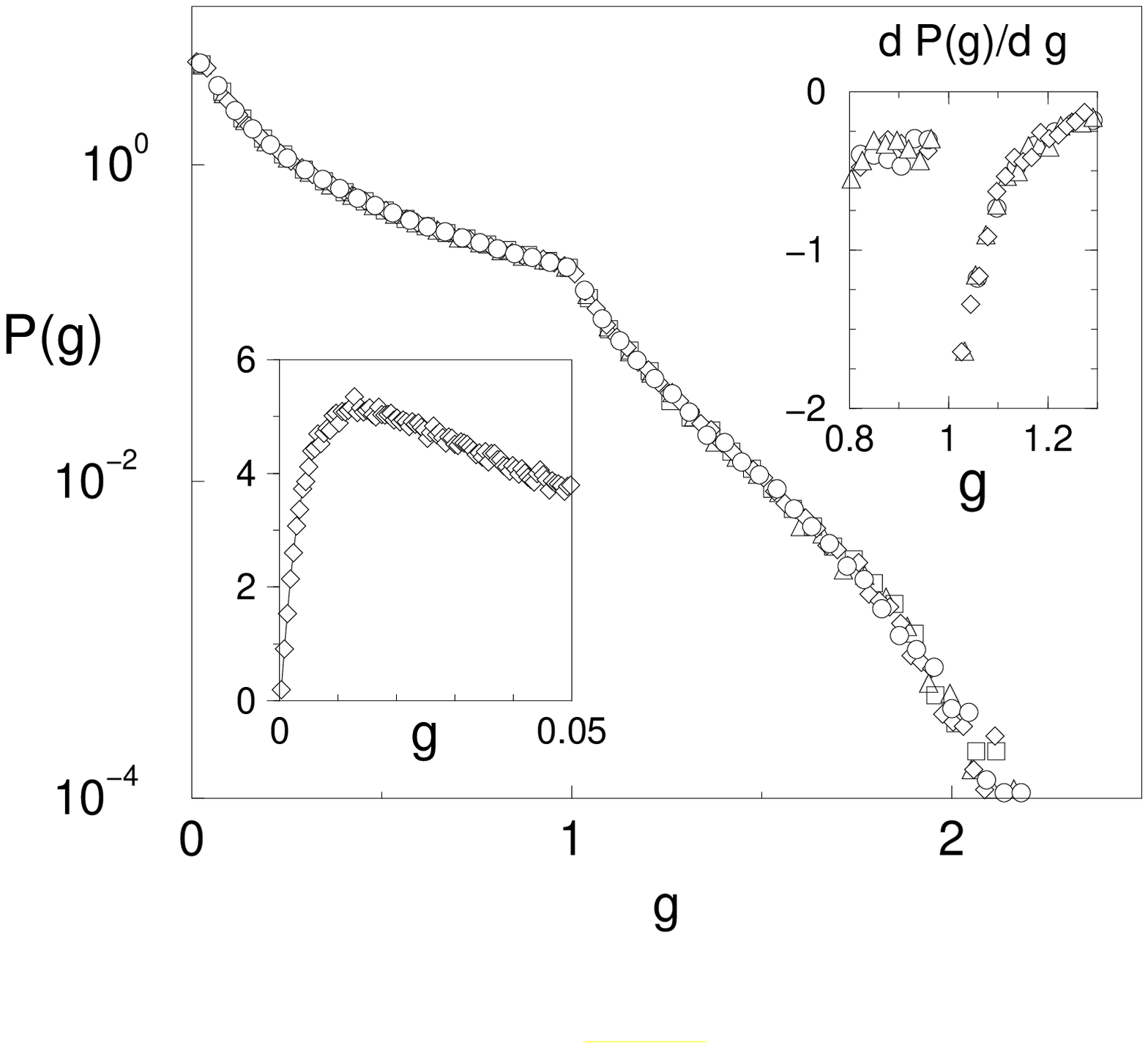}
\includegraphics[width=.24\textwidth]{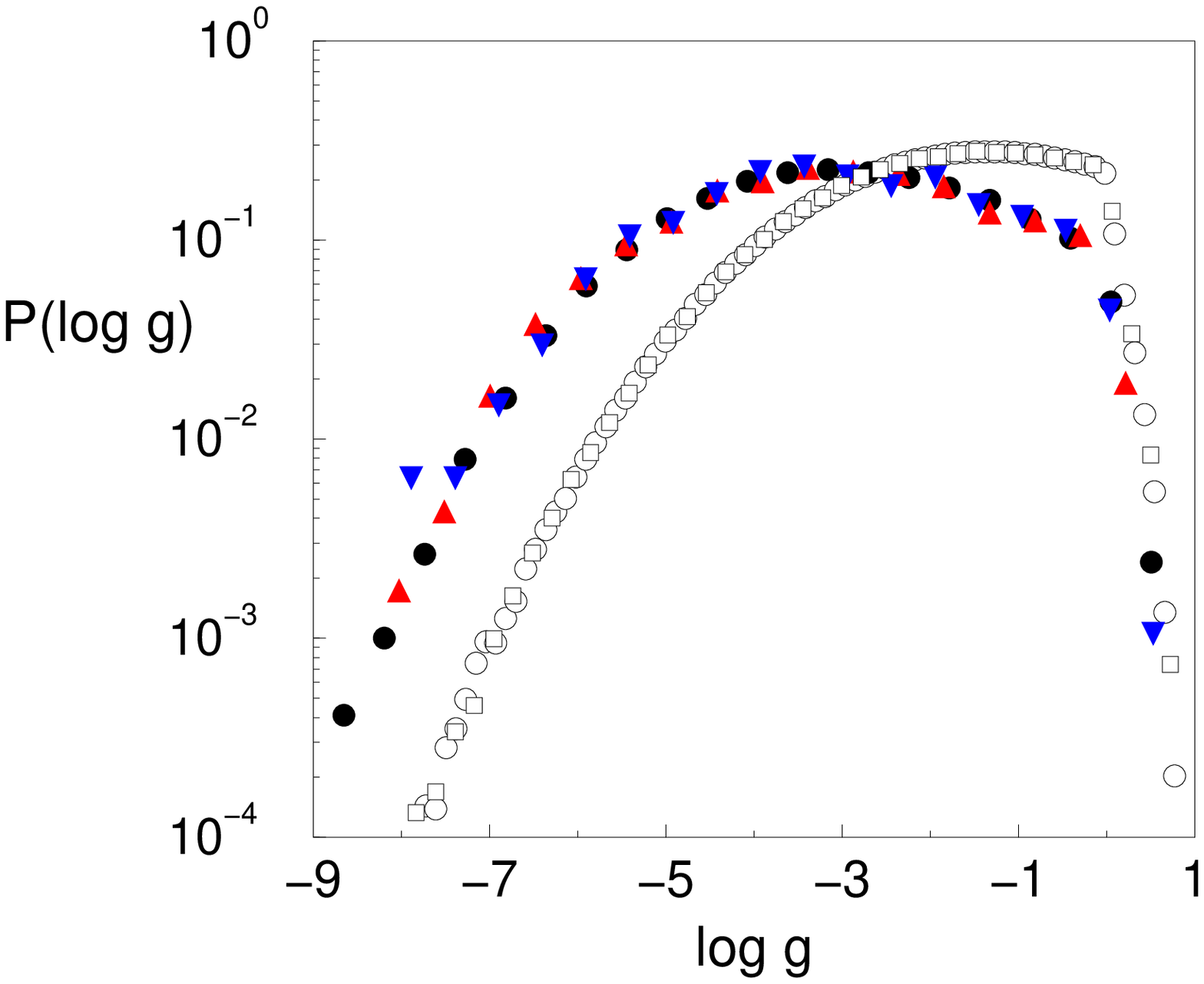}
\end{center}
\caption[]{{\it Left} Critical conductance distribution $P_c(g)$ of the 3D Anderson model, obtained for
statistical ensembles of $N_{\rm stat}=10^6$ samples of the size $10^3 - 18^3$.
$P_c(g)$ is system size independent and decreases faster than exponentially when $g>1$.
The main properties of the critical distribution are
shown in insets:
{\it Left inset} shows in details the small - $g$ behavior and proves that
the distribution decreases to zero as $g\to 0^+$. 
{\it Right inset} shows the discontinuity of the  derivation ${\rm d} P(g)/{\rm d} g$ 
at $g=1$.
{\it Right} figure presents the distribution
$P_c(\ln g)$ at the critical point for the three dimensional  ($W_c=16.5$, {\it open symbols\,}) 
and four dimensional (4D) ($W_c=34.3$, {\it full symbols\,}) systems of various system size.
}
\label{m2}
\end{figure}

\begin{figure}[t!]
\begin{center}
\includegraphics[width=.2\textwidth]{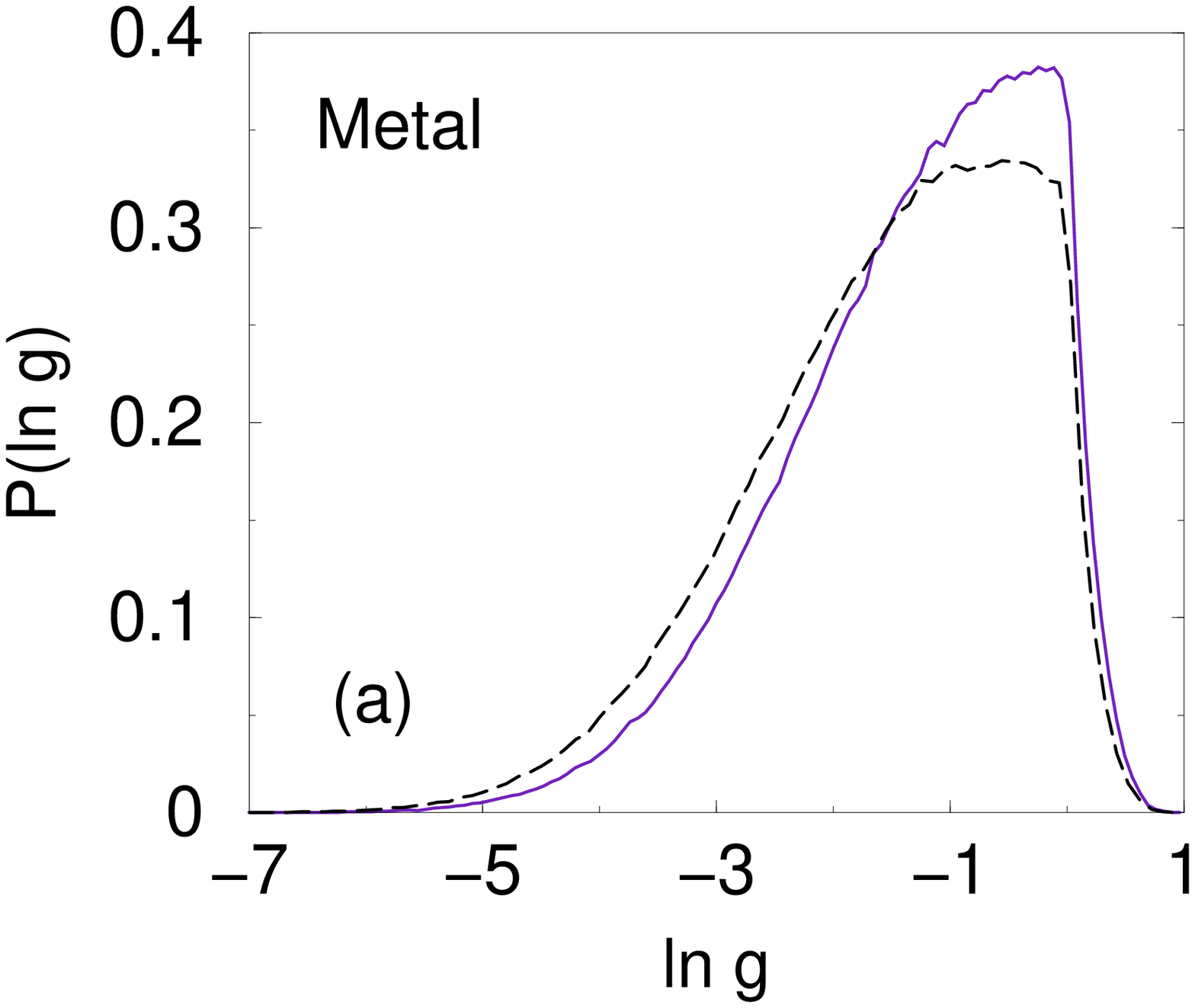}
\includegraphics[width=.2\textwidth]{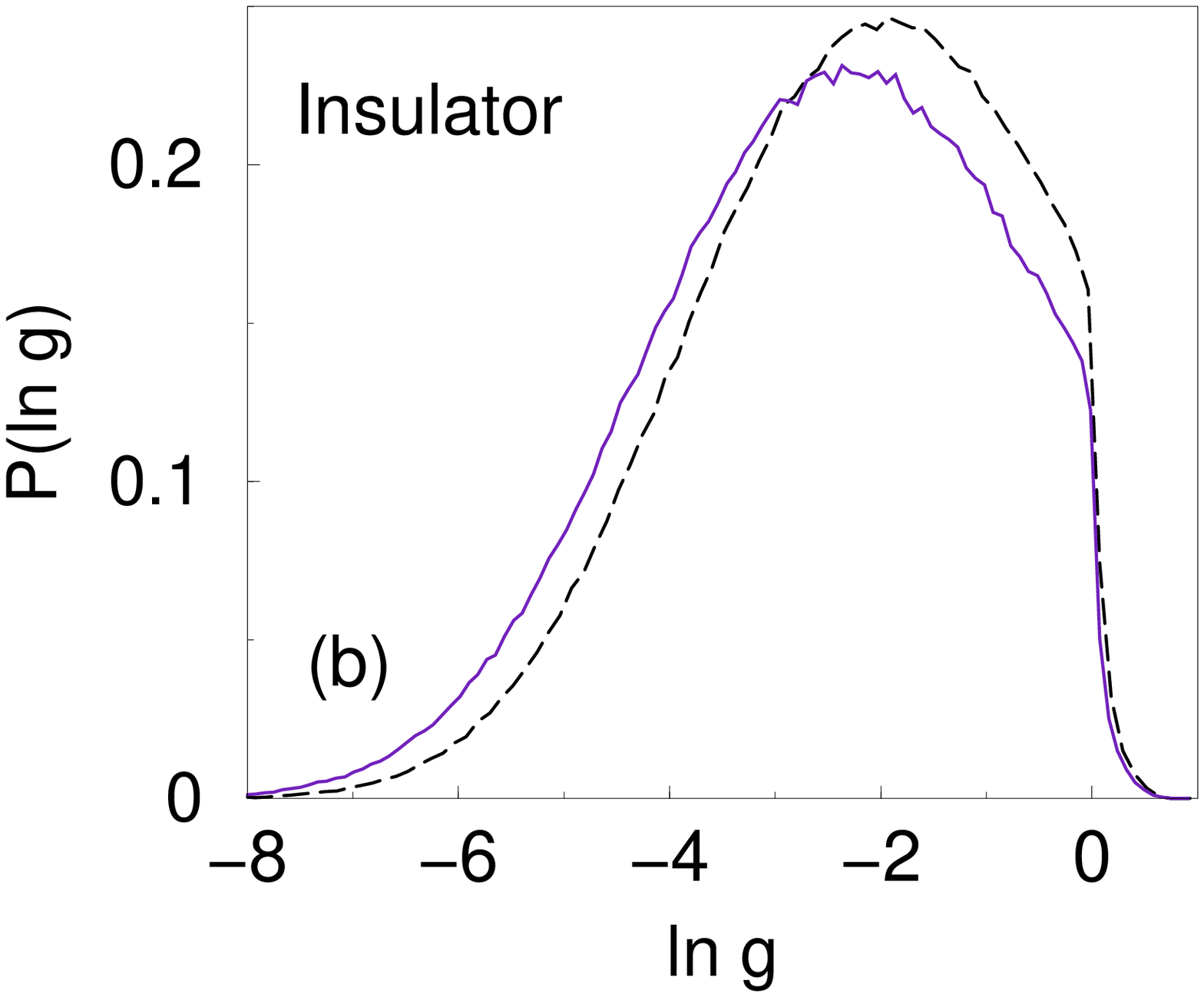}
\end{center}
\caption{Conductance distribution for ensembles of cubes $6^3$ ({\it dashed lines}) and 
$12^3$ ({\it solid line}):  (\textbf{a}) 
The  metallic regime  $W=15.4$  (the critical disorder is $W_c=16.5$).
Distribution $P(g)$ moves toward higher values of the 
conductance as the system size increases and $P(g)$ becomes Gaussian when $L\to\infty$.
(\textbf{b}) 
Insulator $W=17.6$
Mean conductance decreases as the system size increases and the distribution 
of the $\ln g$ becomes  Gaussian 
in the limit $L\to\infty$.
}
\label{m3}
\end{figure}

Suppose that  $g_\alpha$ and $g_\beta$  ($\alpha<\beta$) obey the single parameter scaling. Then
$g_\gamma$ ($\alpha<\gamma<\beta$) scales, too.
Therefore, in contrast to the analysis of the conductance 
cummulants, it is enough to analyze
only a few percentiles. Next, if $g_\alpha$ and $g_\beta$ scale, 
then the difference  $g_\beta-g_\alpha$
scales. Scaling of percentiles assures thus the scaling of the entire conductance 
distribution. Of course, this analysis is not applicable to the limit  $\alpha\to 1$, because available  statistical
ensembles are never  big enough to provide us with sufficient information about the tail of the distribution.

\section{Critical conductance distribution}

At present, we have no analytical description of the critical conductance distribution in 3D systems.
Analytical results were obtained only for the conductance cummulants 
in the dimension $d=2+\epsilon$ close to the lower critical dimension  ($\epsilon\ll 1$)
\cite{AKL}. In spite of the non-universality of higher order conductance cummulants
\xbe
\langle \delta g^n\rangle =\left\{\begin{array}{l} \epsilon^{n-2} ~~~ n<n_0=\epsilon^{-1}\\

                                                   \sim L^{\epsilon n^2-n} ~~~ n>\epsilon^{-1}
					\end{array}\right.
\label{kumulanty}
\xee
the critical distribution $P_c(g)$ was shown  to be universal and $L$-independent in the  limit $L\to\infty$
\cite{CS}. However, theoretical analysis of the form of the critical distribution,
is applicable only in the limit of very small $\epsilon$
\cite{MK-PM}.  

\subsection{The form of the critical conductance distribution}

All what we know about the $P_c(g)$ in 3D is based on the 
numerical data.  In Fig.  \ref{m2} we present $P_c(g)$ for the 3D Anderson model.
Data confirm that the critical conductance distribution is system size independent, as required
\cite{BS}. The shape of  $P_c(g)$ differs considerably from  
the conductance distribution in  the metallic and in the insulating
regimes. To explain the typical properties of the critical conductance distribution, 
we use  our knowledge about statistical properties of parameters  $z$ (\ref{Landauer}).

\begin{figure}[t!]
\begin{center}
\includegraphics[width=.26\textwidth]{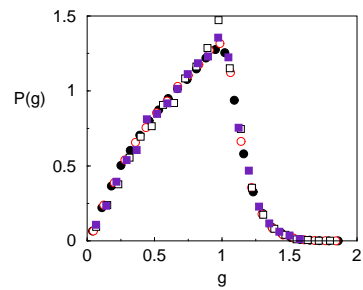}
\end{center}
\caption[]{Comparison of the critical conductance distribution of two different 2D models with
spin-orbit coupling: {\it full symbols} Ando model \cite{Ando-89} {\it open symbols} 
Evangelou-Ziman model \cite{EZ}.  
Squares $40\times 40$ and $80\times 80$ were simulated to prove the system size
independence of the critical distribution. Note also the non-analytic behavior for $g=1$.
}
\label{m4}
\end{figure}

Decrease of $P_c(g)$ to zero when $g\to 0$ is
visible when very large statistical ensembles are studied
(left inset of Fig. \ref{m2}).
Due to Eq. (\ref{Landauer}),
small conductance means that $z_1$ is large.  From numerical data we know that
the distribution  $P(z_1)$ is similar  to Wigner surmises 
and decreases  as $\exp -z_1^2$ for large $z_1$. Consequently,
$\ln P_c(\ln g)$ decreases as $-\ln^2g$ (right figure in Fig. \ref{m2}), and 
guarantees that $\lim_{g\to 0}P_c(g)=0$ \cite{MK-PM,M-PRL}.

Large $g$ behavior of $P_c(g)$ is determined by the chance that many parameters  $z_i$ are  small.   
Statistical analysis of parameters $z_i$ showed that for $i>1$ the
distribution $P(z_i)$ is Gaussian with 
mean value $\langle z_i\rangle\propto \sqrt{i}$ and
var $z_i\propto \langle z_i\rangle^{-1}$
\cite{M-PRL}. The probability to find a sample with small value of  
higher $z_i$  is therefore very small: $P_c(g\approx i)$ decreases as $\exp-i^{3/2}$
\cite{M-PRL}.
Fig. \ref{m2} indeed shows  very fast decrease of the probability $P_c(g)$ for  
$g>1$. The chance to have $g>1$ is only $3\%$.
Probability to find  large values of  $g$ drastically decreases: 
we found that in the ensemble of $10^7$ samples ($L=10$) 
only 470 samples have $g>2$  and only one sample has  $g>3$ \cite{RMS-01a}. 
The analysis of  the contribution of the first two channels is  therefore 
sufficient for the understanding of the qualitative properties  of the critical distribution.

\begin{figure}[t!]
\begin{center}
\includegraphics[width=.2\textwidth]{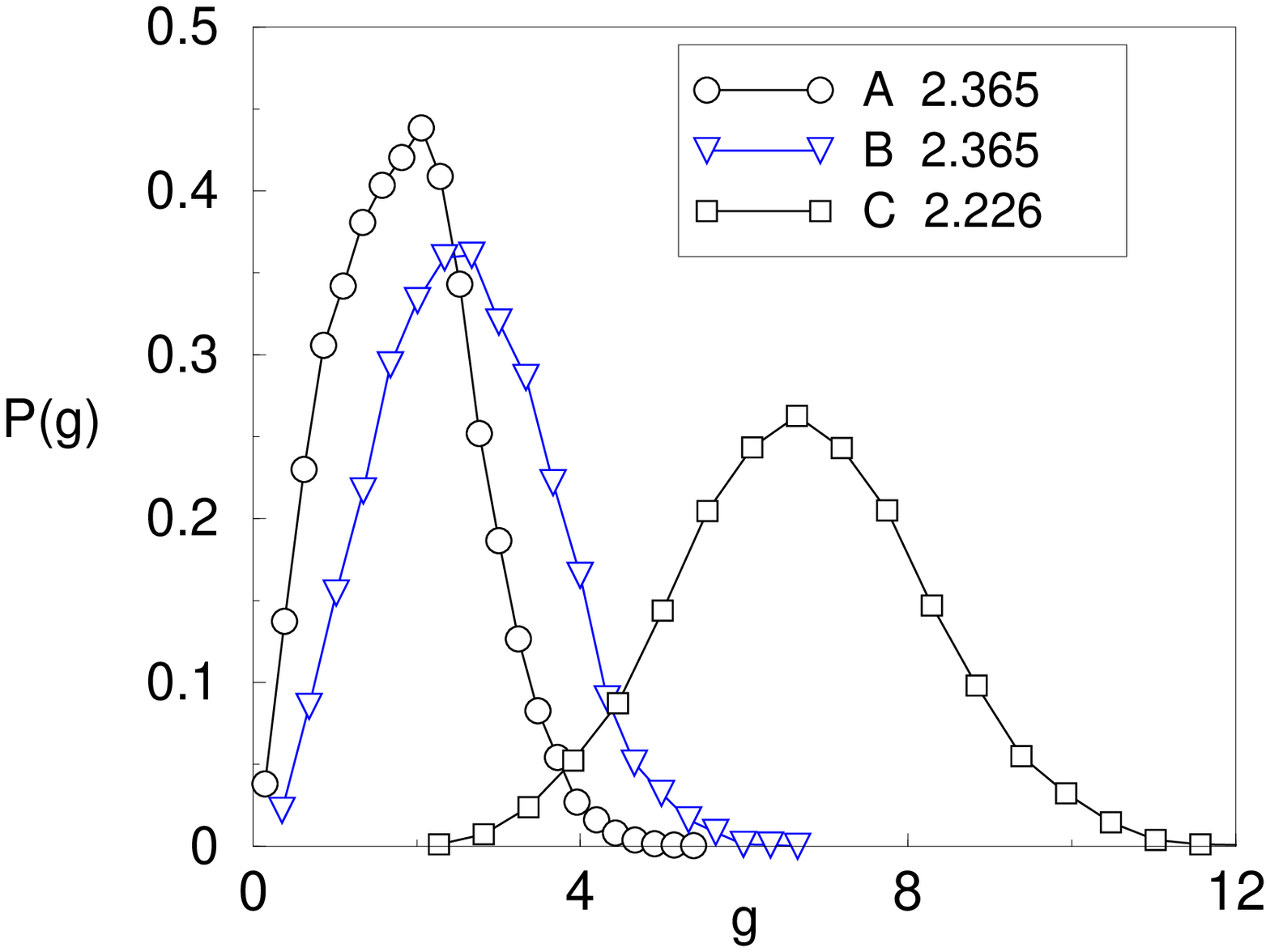}
\includegraphics[width=.2\textwidth]{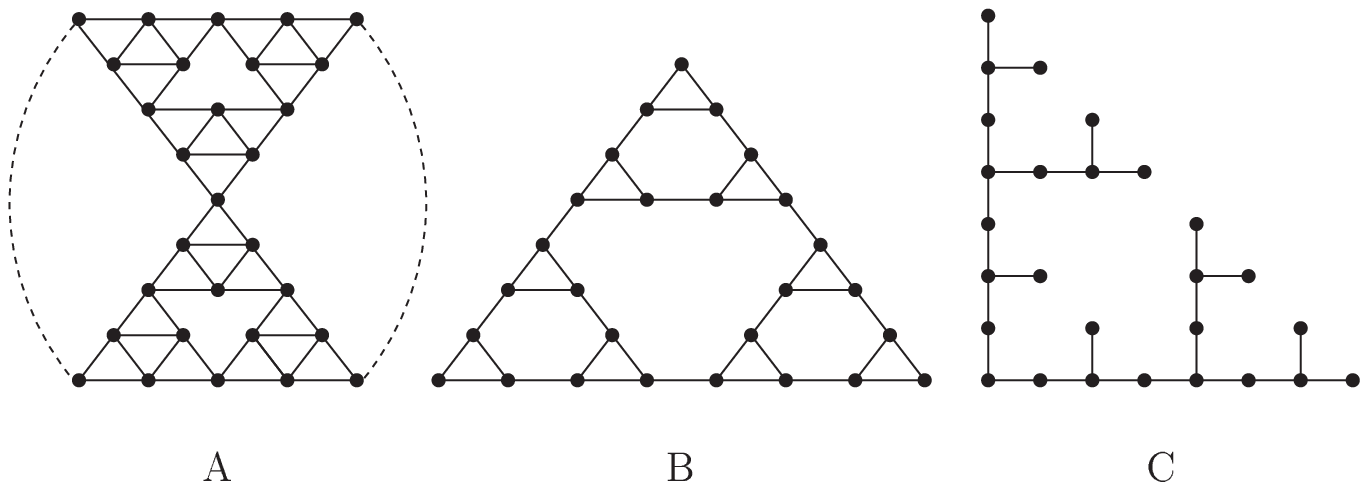}
\end{center}
\caption[]{Critical conductance distribution for  bifractal lattices. 
Legend gives spectral dimension.
5th generation of fractals was used to calculate the distribution. Bifractal is 
linear along the $z$ axis and fractal in a plane perpendicular to  $z$.
{\it Right} figure shows the structure of the fractal (the 3rd generation) 
cross section.  All fractals have the same  fractal dimension $d_{\rm f}=\ln 3/\ln2$.
Spectral dimension $d_{\rm s}=\ln 9/\ln 5$ for fractals A and B, and $=\ln 9/\ln 6$ for fractal C.
Note that although  
bifractals A and B have the same both fractal and spectral dimensions, 
they posses different critical conductance distribution. $P_c(g)$ depends on the lattice
topology \cite{TM-PRB}}
\label{m5}
\end{figure}

Numerical data also show that the critical distribution is non-analytical
at  $g=1$.  Right inset of Fig. \ref{m2} shows the discontinuity of  the first
derivative ${\rm d} P_c(g)/{\rm d} g$. 
The same   non-analyticity was found  in  the 4D systems (right figure
\ref{m2}), in the unitary \cite{SO-97} and symplectic (Fig. \ref{m5}) systems, and also in the
weakly disordered quasi-1D systems \cite{MuttW-99,M-PRB02a}.

Present description of the critical distribution is based on the analysis of 
statistical properties of parameters $z$. It is applicable to any system, for which
the mean values of parameters $z$ are of the order of unity. Then only a few (two, or three) channels
contribute to the conductance. 
This  analysis is, however, not applicable to systems close to the lower 
critical dimension
$d=2+\epsilon$. Here, the mean conductance $\langle g\rangle\sim\epsilon^{-1}$, 
which means that number of channels which contribute to the conductance, is large, $\sim\epsilon^{-1}$. 
It is therefore no surprise that the
critical conductance distribution found in \cite{CS} differs from that shown in
fig. \ref{m2}.

\subsection{Conductance distribution in the critical regime}

Fig. \ref{m3} shows that  the  typical properties of the
critical conductance distribution
hold also for $P(g)$  in the neighbor of the
critical point. This is because the statistical properties of parameters $z$ depend 
continuously
on the disorder in the critical regime. Critical properties of 
$P(g)$ survive until the system size $L$ exceeds the correlation length $\xi$.
Only when  $\xi\ll L$
the distribution  typical for the  metallic or the  insulating regime can be observed.

\subsection{Universality}

Single parameter scaling theory of localization supposes that the critical conductance
distribution is universal. Its form does not depend on the microscopic
details of the model.
As an example, we present in Fig. \ref{m4} 
the critical conductance distribution for two 2D models with spin-orbit scattering: 
Evangelou-Ziman model \cite{EZ}  and Ando model \cite{Ando-89}.
In spite of the different 
microscopic definition of both models, $P_c(g)$ is universal
\cite{M-EL}.  Universality of $P_c(g)$ with respect to various distributions of random energies $\varepsilon_n$ was 
confirmed in \cite{M-EL} and \cite{SO-01}. 

As was shown already in  fig. \ref{m2},
the shape of $P_c(g)$ depends on the dimension of the model. Ref. \cite{SO-97}
confirmed that also the physical
symmetry influences the  form of $P_c(g)$.
Less expectable was the observation \cite{BHMMcK,SOK-00}
that  $P_c(g)$ 
as well as the spectral statistics \cite{BMP}
depend  on the {\it boundary conditions} in the transversal direction.
Nevertheless, this is consistent with the original definition of the conductance
as a measure of the sensitivity of the energy spectrum of the system
to the change of the boundary conditions \cite{LT}.

For completeness, let us note that the critical conductance distribution depends also on other parameters of
the model: lattice topology \cite{TM-PRB}, anizotropy \cite{RMS-01b} and, of course, on the length of the 
system. We believe that these non-universalities could be compensated by the change of another model  parameter
(see, for instance \cite{RMS-01b}, where the anizotropy is compensated by the length of the system).

\subsection {Dimension dependence}

Right figure in Fig.  \ref{m2} compares the critical conductance
 distribution for 3D and 4D cubes.
As supposed, the maximum of  $P_c(g)$ for 4D is shifted toward smaller conductances, because
the critical disorder increases as the spatial dimension increases and
higher disorder means lower mean conductance \cite{Wegner-76,CS}.
Qualitatively, however, both distributions are  very similar:
$P_c(\ln g)$ decreases as $\exp[-\ln^2 g]$ for $\ln g\to -\infty$  and possesses
the non-analyticity 
at $\ln g=0$. This similarity is not surprising, 
because the form of the distribution
is determined mostly by the statistics of  $z_1$ and $z_2$,
which are qualitatively similar in 3D and 4D
\cite{M-PRL}. 

Surprisingly, the relation (\ref{kumulanty})  seems to hold also
for $\epsilon=1$ and 2, at least for the first two cummulants.
We obtained numerically that $\langle g\rangle=0.285$ 
for 3D and 0.135
for 4D, so that 
$\langle g\rangle_{\rm 3D} \approx 2\langle g\rangle_{\rm 4D}$.
For the second cummulants we found
${\rm var}~g_{\rm 3D}\approx {\rm var} g_{\rm 4D}\approx 0.17$ \cite{TM-PRB}.

\smallskip

More interesting is the investigation of the $P_c(g)$ in the systems
of dimension $2+\epsilon$
\cite{TM-PRB}. As we are not able to create the $d$-dimensional hyper-cubes 
with non-integer $d$ in computers, we simulated the transport on bifractal latices 
\cite{SG,TM-PRB}. Bifractals
are linear along the propagation direction and possess the fractal lattice in the cross section
(fig. \ref{m5}).
We proved that the critical  exponent $\nu$ depends 
only on the {\it spectral} dimension of the lattice. Mean 
conductance, var $g$   and the critical distribution $P_c(g)$  depend, however, on the lattice topology.
For instance, fig. \ref{m5} shows that
bifractals A and B have the same spectral dimension, but different critical distribution.
This is the reason why  obtained data can not be used for the verification of relations 
(\ref{kumulanty}). 

In Fig. \ref{m5} we present $P_c(g)$ for  three different 
bifractals. As expected, $\langle g\rangle$ increases
and the distribution converges to Gaussian when $\epsilon\to 0$. However, we found neither the
$\delta$ - function   peak at $g=0$
nor  the power-law tail of the distribution
for $g\gg \langle g\rangle$, predicted by the theory \cite{CS}.

\section{Conductance distribution in non-critical regime}

Although we have no analytical theory of the conductance statistics in the critical regime, we can learn some
typical properties of the conductance distribution from the analysis of the 
quasi-1D weakly disordered systems \cite{MuttW-99}. 
Starting from the Dorokhov - Mello-Pereyra-Kumar equation \cite{DMPK} for the
probability distribution of parameters $z$,
the conductance distribution 
$P(g)$ can be calculated.
$P(g)$ depends on the length of the system.

\begin{figure}[t!]
\begin{center}
\includegraphics[width=.2\textwidth]{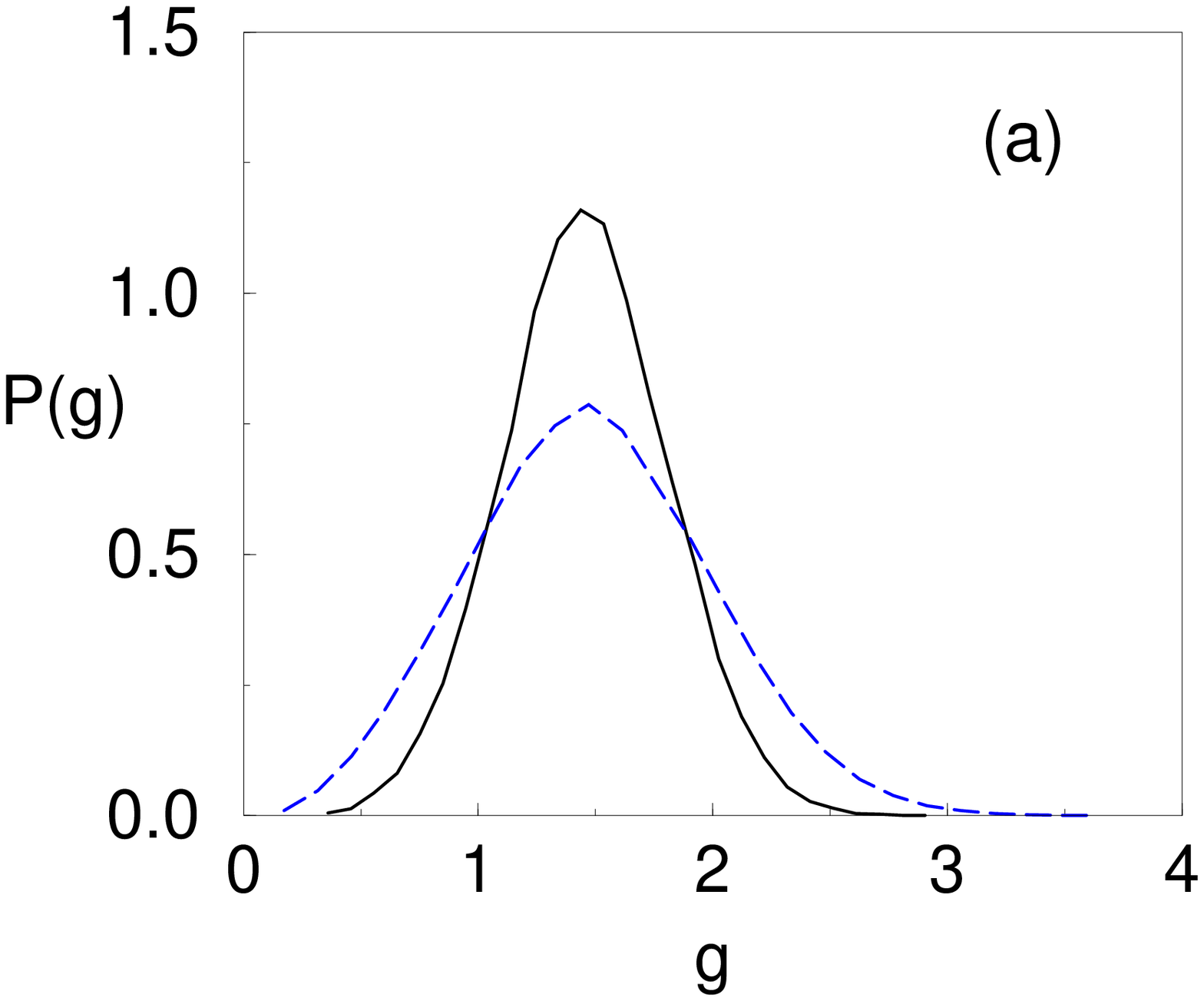}
\includegraphics[width=.2\textwidth]{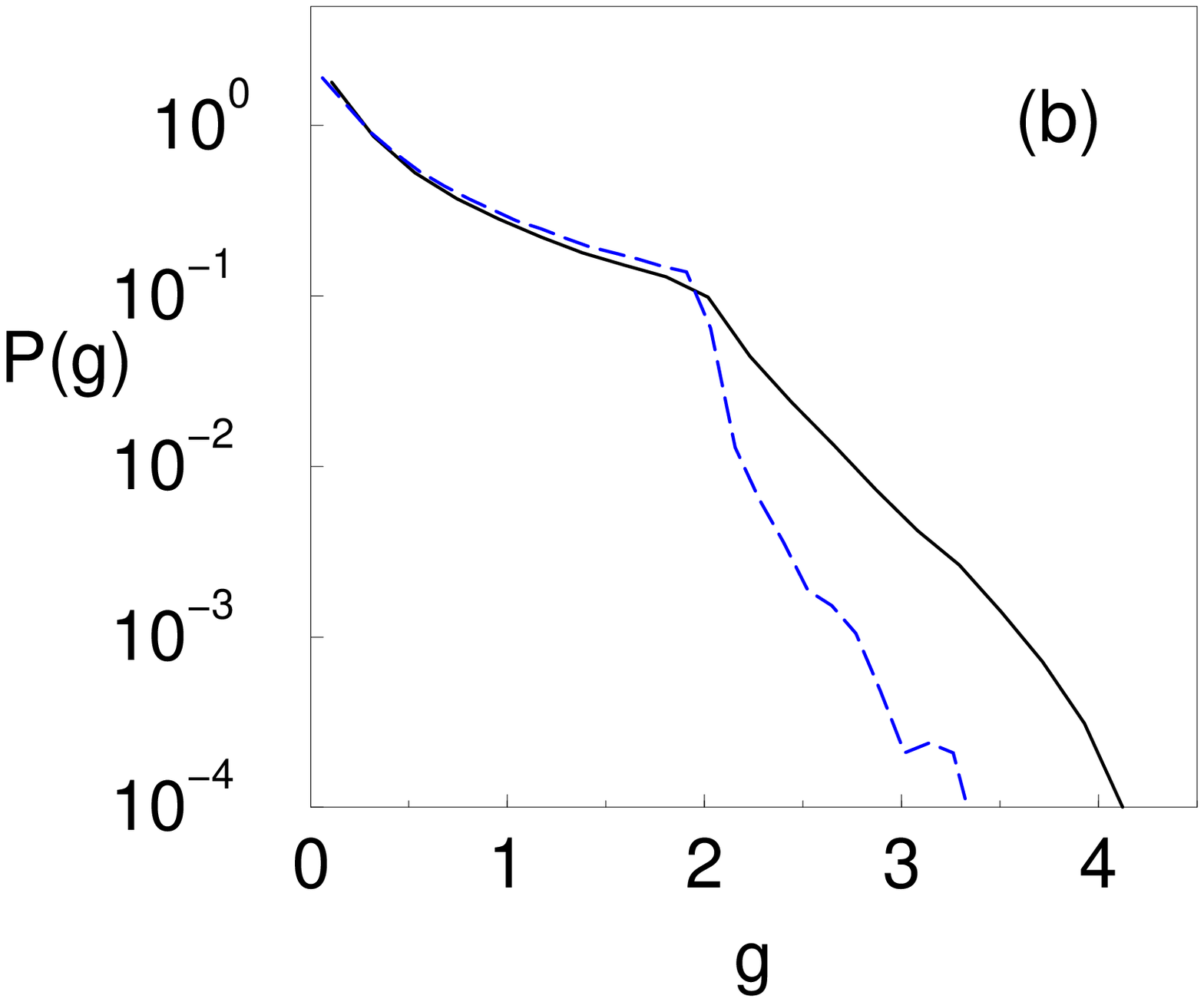}
\includegraphics[width=.2\textwidth]{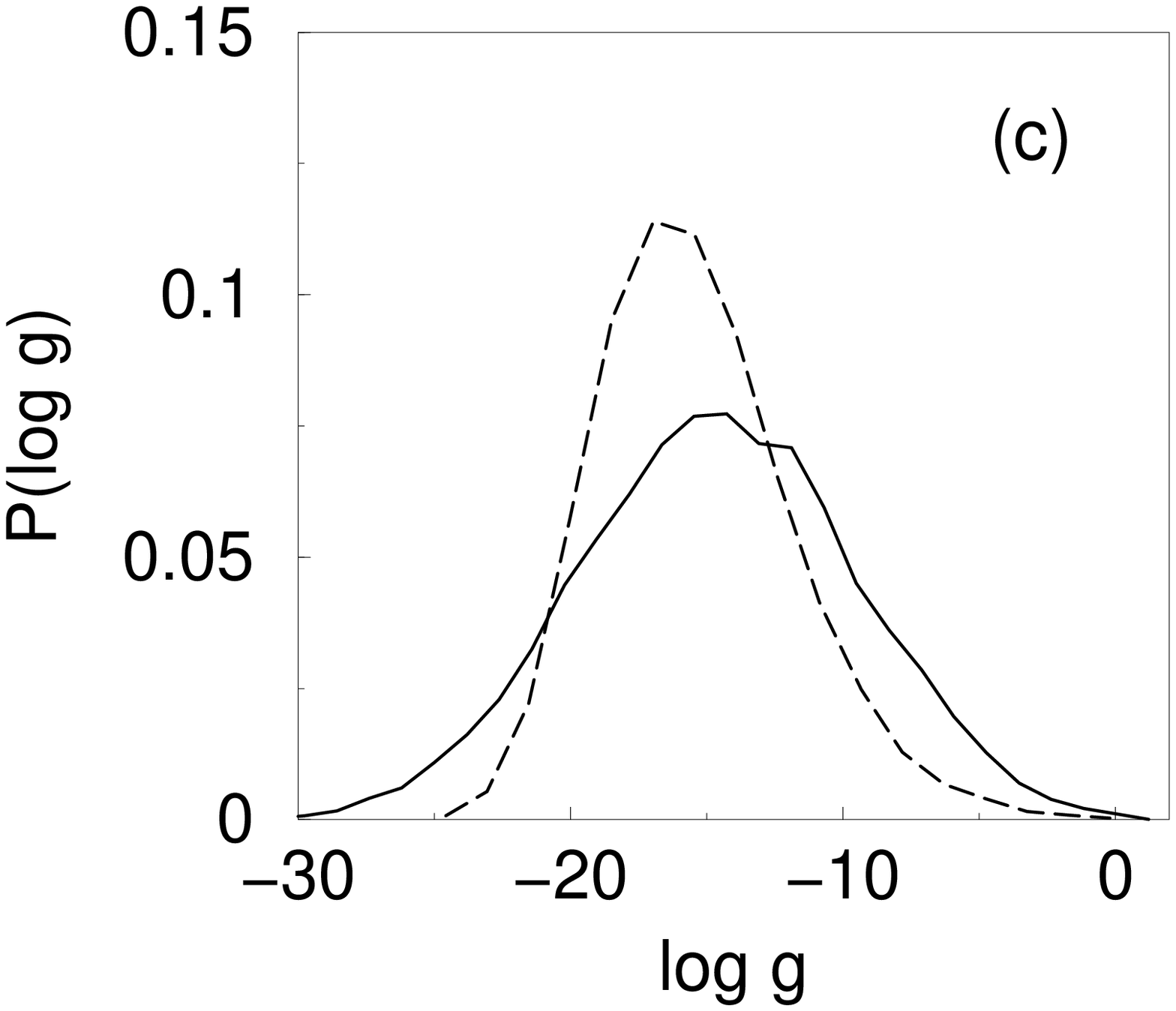}
\end{center}
\caption[]{Comparison of the conductance distribution of the 3D ({\sl dashed lines})
and  quasi-1D 
({\it solid line}).
{\bf (a)} Metallic ($\langle g\rangle>1$)  {\bf (b)} critical 
($\langle g\rangle\approx 1$) and {\bf (c)} localized ($\langle g\rangle\ll 1$) 
regimes are shown. 
In the quasi-1D systems, the strength of the disorder is fixed to $W=4$ and the length of the system is 
tuned to obtain the same
mean conductance as in the 3D system.}
\label{m6}
\end{figure}

\begin{figure}[t!]
\begin{center}
\includegraphics[width=.22\textwidth]{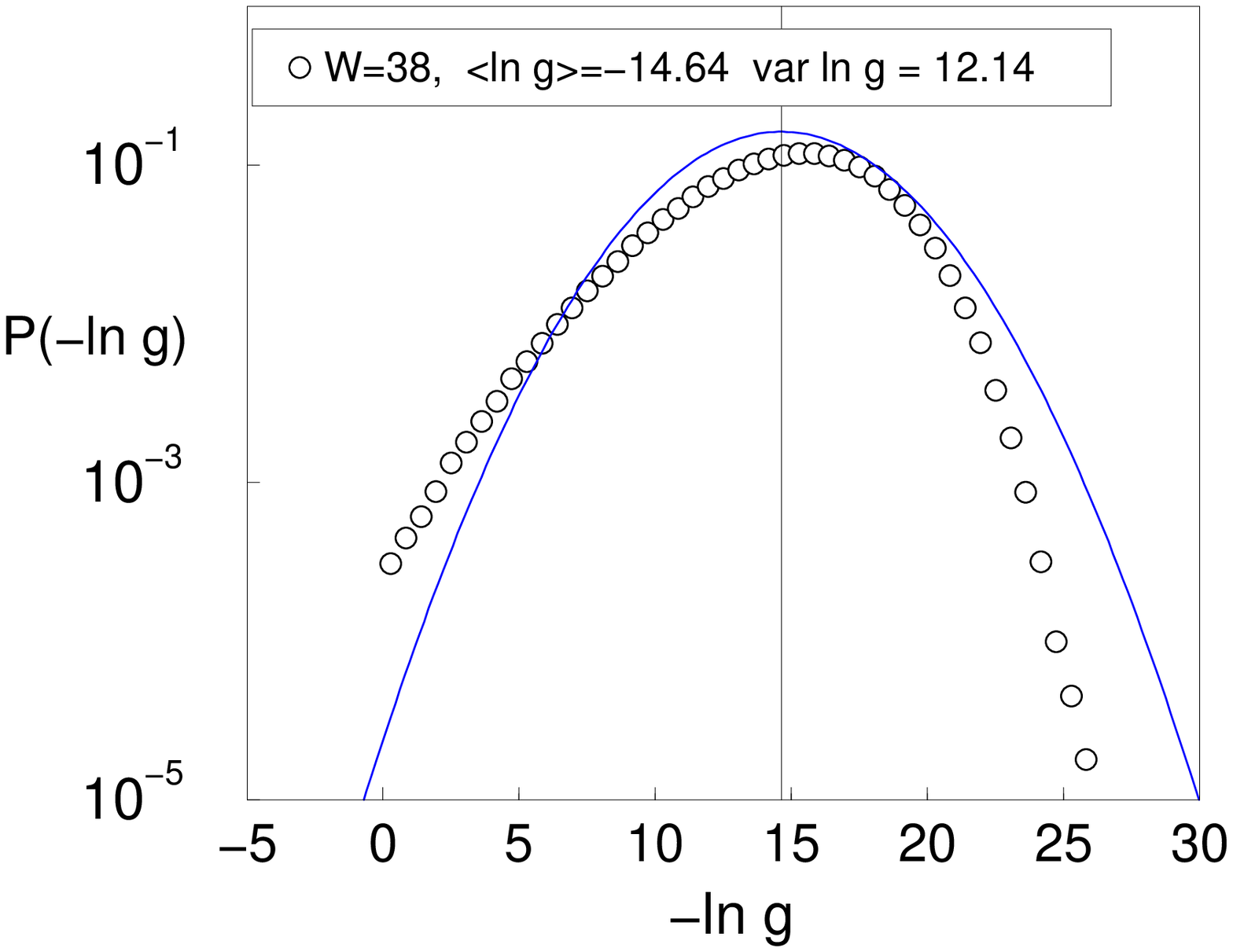}
\includegraphics[width=.22\textwidth]{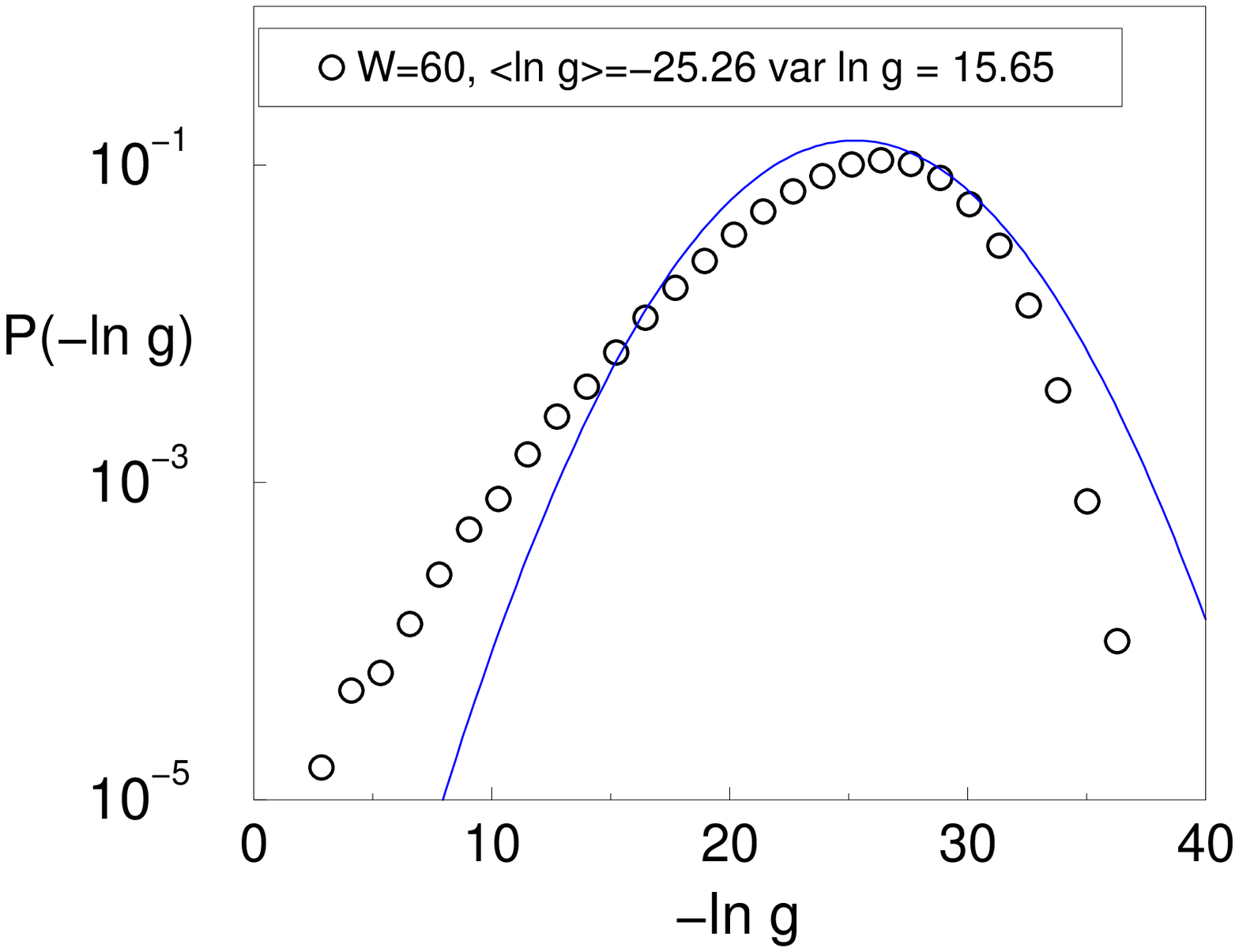}
\end{center}
\caption[]{Distribution of  
$-\ln g$ for strongly localized regime.
Note that the distribution is not symmetric. It possesses the long tail
for small values of the argument, and decreases much faster on the opposite side.
{\it Solid} lines are a  Gaussian distribution  with the mean and variance 
found from numerical data. Size of the cubes is
$L=10$. {\it Left}: $W=38$. Note that the sharp decrease of the distribution at $\ln g=0$, discussed in Sect.
3.1, is still present \cite{MuttW-99}.  {\it Right}:  $W=60$. 
}
\label{m7}
\end{figure}

\noindent For short samples, $P(g)$ is Gaussian
\cite{Pnato} (see also fig. \ref{m6}a).
When  the length of the system increases, the conductance decreases and
the system passes from the metallic regime into  the localized one.
For the intermediate system length, where $\langle g\rangle\approx 1$, $P(g)$ 
is expected to be {\it qualitatively} 
similar to the critical conductance distribution
\cite{MuttW-99}. This is shown in
Fig \ref{m6}, where we compare the conductance distribution of quasi-1D
and 3D systems of the same mean conductance.

Fig. \ref{m6} also shows that the analogy between 3D and quasi-1D systems is not exact.
To understand {\it quantitative} differences between the quasi-1D and 3D systems, 
we analyzed the spectrum of parameters $z$. 
In the metallic regime, the difference between 3D and quasi-1D systems
is only in the  value of var $g$ \cite{LSF}.  $P(g)$ is Gaussian and
the spectrum of parameters $z$ is linear $\langle z_i\rangle\propto i$ \cite{Imry,Pnato}
independently on the dimension of the system \cite{M-JPCM-95}.
In contrast to the metallic regime,
the spectrum of $z$ becomes dimension dependent
in the critical region.
Both quasi-1D \cite{MH-JPCM,M-JPCM-95} and 3D-dimensional 
\cite{M-PRB02a} numerical studies confirmed that at the critical point
\xbe
\langle z_i\rangle^{d-1}\propto i
\label{dd}
\xee
($d>2$). 
Owing to (\ref{dd}), the difference $\Delta=\langle z_2-z_1\rangle$ is smaller that
$\langle z_1\rangle$ in 3D, while it equals to $\langle z_1\rangle$ in the quasi-1D system.
The contribution of the second channel is therefore more important in 3D than in the quasi-1D.
This explains longer tail of the distribution in the 3D system (fig. \ref{m6}b).

It is commonly believed that the distribution $P(\ln g)$ is
Gaussian in the insulating phase, independently on the dimension of the system.
This is,however, not true. The
spectrum of $z$ depends namely on the dimension of the system also in the localized regime. 
For the 3D systems
it  was proved numerically \cite{M-PRB02a,M-JPCM-95}
that the difference $\Delta=\langle z_2-z_1\rangle$ is {\it constant},
independent on the disorder and on  the  system size. Therefore
the second channel influences always the statistics of the conductance.
More than  its contribution to the  value of the conductance it is important
that constant  value of $\Delta$ prevents
the distribution $P(z_1)$ to develop into the Gaussian form. 
While the values 
$z_1\ll \langle z_1\rangle$ are still possible, the probability to find systems
with much higher values
$z_1\gg \langle z_1\rangle$ is  strongly suppressed.
The distribution $P(z_1)$ is therefore not symmetric. 
The same is true for the distribution $P(-\ln g)$ (Fig. \ref{m7})
which  possesses a long tail for
small values of $|\ln g|$   and decreases much faster than Gaussian for
$|\ln g|\to\infty$. 

Note that in weakly disordered quasi-1D systems $\Delta\sim \langle z_1\rangle$.
The distance  $\langle z_2-z_1\rangle$ is much larger than the width of the distribution $P(z_1)$.
Higher channels therefore do not influence the distribution of $z_1$ and $P\ln g$  is Gaussian.

\section{Conclusion}

We reviewed  recent  progress in numerical studies of the statistics of the conductance
in the critical regime. 
Numerical analysis confirms  that
the conductance distribution in 3D Anderson model obeys 
single parameter scaling. 
Analysis of the statistics of the eigenvalues of the transmission matrix
enables us to understand the main features of the
conductance distribution in the critical regime.
Although we  still have no 
analytical  description of  the conductance statistics in the critical regime,
we hope that results of numerical experiments will inspire  theoreticians to formulate
the general analytical theory of the Anderson transition.

\end{document}